\documentclass[12pt]{article}
\usepackage{latexsym}

\newcommand{\be}{\begin{equation}}
\newcommand{\ee}{\end{equation}}
\newcommand{\bea}{\begin{eqnarray}}
\newcommand{\eea}{\end{eqnarray}}

\newcommand{\noi}{\noindent}
\newcommand{\nn}{\nonumber}

\newcommand{\lesssim}{ {\
\lower-1.2pt\vbox{\hbox{\rlap{$<$}\lower5pt\vbox{\hbox{$\sim$}}}}\
} }
\newcommand{\gtrsim}{ {\
\lower-1.2pt\vbox{\hbox{\rlap{$>$}\lower5pt\vbox{\hbox{$\sim$}}}}\
} }

\begin{document}

\begin{titlepage}

%\begin{flushright} \\
%\today
%\end{flushright}
%\vspace*{1.5cm}
\begin{center}
{\Large \bf On the use of the Operator Product Expansion \\
[.5cm]to
constrain the hadron spectrum.}\\[3.0cm]

{\bf Maarten Golterman}$^a$ and {\bf Santiago Peris}$^b$\\[1cm]

$^a$Department of Physics and Astronomy, San Francisco State University\\
1600 Holloway Ave, San Francisco, CA 94132, USA\\
e-mail: maarten@quark.sfsu.edu\\[0.5cm]
$^b$Grup de F{\'\i}sica Te{\`o}rica and IFAE\\ Universitat
Aut{\`o}noma
de Barcelona, 08193 Barcelona, Spain\\
e-mail: peris@ifae.es

\end{center}

\vspace*{1.0cm}

\begin{abstract}

We call into question a recently proposed idea on how the
short-distance behavior of QCD correlation functions constrains
hadronic states to become degenerate parity eigenstates as one
goes up in the spectrum.  In particular, we point out that the sum
rules which have been proposed in this context are in general
regulator dependent, and thus ill-defined.

\end{abstract}

\end{titlepage}

%\section{Introduction}
%\lbl{sec:int}

\noi

In QCD with an infinite number of colors, meson resonances are
infinitely narrow and, consequently, can be precisely identified
as quantum states.  Furthermore, because of asymptotic freedom
and confinement, the number of meson resonances in each channel
is infinite \cite{largeN}. Given an operator with the appropriate
quantum numbers (such as a (partially) conserved current), each
state is characterized by two parameters, its mass, and the
amplitude of the operator between this state and the vacuum.  In
the absence of a solution to large-$N_c$ QCD, it is interesting
to see how much the values of these parameters can be constrained
on the basis of large-$N_c$, chiral symmetry, and the Operator
Product Expansion (OPE).

An interesting attempt to do this was presented in
Ref.~\cite{BeanePRD}. The main result of this analysis is that
vector and axial-vector mesons would have to pair up as one goes
up in the spectrum  so that chiral symmetry is restored at higher
energies. In particular, two sum rules similar to those derived
by Weinberg long ago \cite{Weinberg} were proposed, and applied to
various models of the meson spectrum in the (non-singlet) vector
and axial-vector channels. These sum rules were derived in two
complementary ways. The first derivation is based only on very
general properties of the OPE, and the second one uses only
chiral symmetry and null-plane charges.\footnote{See also
Ref.~\cite{BeaneJPG}}

These results of Ref.~\cite{BeanePRD} have been used to infer the
spectrum of hybrids \cite{BeanePLB}, to ``test" models (such as
that of Ref.~\cite{Golterman}), and to argue for the phenomenon
of  chiral symmetry restoration for highly excited states in the
hadronic spectrum \cite{CohenIJMP,CohenPRD,Glozman}.  Since these
claims are rather far-reaching, we consider it indispensable to
have a closer look at the theoretical foundations on which the
analysis of Ref.~\cite{BeanePRD} is based.

Let us start by defining the (covariantly time-ordered) vector
and axial-vector two-point functions in the chiral limit as \be
\label{correlator} \Pi_{V,A}^{\mu\nu}=\ i\,\int
d^4x\,e^{iqx}\langle 0|T(J_{V,A}^\mu(x) J_{V,A}^{\dag \nu}(0))|0\rangle
= \left(q_{\mu} q_{\nu} - g_{\mu\nu} q^2 \right)\Pi_{V,A}(q^2) \
, \ee with $J_{V}^\mu(x) = {\overline d}(x)\gamma^\mu u(x)$ and
$J_A^\mu(x) = {\overline d}(x)\gamma^\mu \gamma^5 u(x)$. Both
functions $\Pi_{V,A}(Q^2\equiv -q^2)$ satisfy the dispersion
relation (up to one subtraction) \be \label{dispersion}
\Pi_{V,A}(Q^2)= \int_0^{\infty} \frac{dt}{t+Q^2}\ \frac{1}{\pi}
{\rm Im}\,\Pi_{V,A}(t)\ . \ee In large-$N_c$ QCD one finds that
 \be \label{Imvector}
\frac{1}{\pi} {\rm Im}\,\Pi_V(t)= 2 \sum_{n=1}^{N_{V}} F_V^2(n)
\delta(t-M^2_V(n))\ , \ee \be \label{Imaxial} \frac{1}{\pi} {\rm
Im}\,\Pi_A(t)=  2 \sum_{n=1}^{N_A} F_A^2(n) \delta(t-M^2_A(n))\ ,
\ee where $F_{V,A}(n)$ and $M_{V,A}(n)$ are the two parameters
for each resonance $n$ in each channel ($V$ or $A$) mentioned
above,
with $n=1$ in the axial channel corresponding to the pion.%
\footnote{Therefore, $ F_A^2(n=1)\equiv F_{\pi}^{2}=(93\
\mathrm{MeV})^2$ and $M_A^2(n=1) \equiv M_{\pi}^{2}=0$ in the
chiral limit, to which we will restrict ourselves in this paper.}
It follows that
\begin{eqnarray}\label{PiV}
  \Pi_{V}(Q^2)&=&\sum_{n=1}^{N_V}
  \frac{F_V^2(n)}{Q^2+M_V^2(n)}\ ,\\\label{PiA}
\Pi_{A}(Q^2)&=& \sum_{n=1}^{N_A} \frac{F_A^2(n)}{Q^2+M_A^2(n)}\ ,
\end{eqnarray}
where it is understood that $N_{V,A}$ are to be taken to infinity
after physical observables have been calculated, with  $Q^2\ll
M^2_{V}(N_V), M^2_{A}(N_A)$. We thus emphasize the need for an
ultraviolet cutoff in these sums.

At large values of $Q^2$, $\Pi_{V,A}$ in
Eqs.~(\ref{PiV},\ref{PiA}) have to reproduce the parton-model
logarithm. This follows easily with the Euler-Maclaurin summation
formula, which, applied to our case, reads
\begin{eqnarray}\label{Euler}
\Pi_{V,A}(Q^2)&=&\ \int_{0}^{N_{V,A}+1} dn
\frac{F^2_{V,A}(n)}{Q^2+M^2_{V,A}(n)} \nn \\
&&\quad - \frac{1}{2}\left\{
\frac{F^2_{V,A}(0)}{Q^2+M^2_{V,A}(0)}+
\frac{F^2_{V,A}(N_{V,A}+1)}{Q^2+M^2_{V,A}(N_{V,A}+1)}\right\}+...
\qquad .
\end{eqnarray}
This indeed reproduces a ``$\log{Q^2}$'' when
$Q^2\rightarrow\infty$ (as long as we take $N_{V,A}$ to infinity
first) if one assumes that\footnote{Note that in general
$F^2_{V,A}(n)\ \sim \frac{d}{dn} M^2_{V,A}(n)$ would do just as
well, so Eq.~(\ref{asympt}) is not the only possibility.}
\begin{equation}\label{asympt}
M_{V,A}^2(n)\sim n\ ,\ \ \ \ F_{V,A}^2(n)\sim F_{V,A}^2 \quad ,
\end{equation}
for $n\rightarrow\infty$, with $F^2_{V,A}$ constants independent
of $n$. This assumption is for example in accord with Regge
Theory in which the ``daughter trajectories" are given by
\begin{equation}\label{Regge}
  M^2_{V,A}(n)= \Lambda^2_{V,A}n+\mathrm{constant}\ .
\end{equation}
As a matter of fact, the behavior of Eq.~(\ref{asympt}) is what is
obtained in two dimensions, for which large-$N_c$ QCD can be
solved \cite{Callan}.  It is reasonable to expect that
Eq.~(\ref{asympt}) maybe true also in four dimensions.

We know from perturbation theory that the coefficient of the
logarithm in $\Pi_V$ and $\Pi_A$ is the same. This leads to the
further condition
\begin{equation}\label{PT}
  \frac{F_V^2}{\Lambda_V^2} = \frac{F_A^2}{\Lambda_A^2}= \frac{N_c}{24 \pi^2} \ .
\end{equation}
Furthermore, we know that
$\Pi_{LR}(Q^2)=\Pi_{V}(Q^2)-\Pi_{A}(Q^2)$ should vanish for
$Q^2\to\infty$, in order to avoid conflict with chiral symmetry
and asymptotic freedom.  It follows directly from Eq.~(7) that
this is only the case if the limits $N_{V,A}\to\infty$ are taken
such that
\begin{equation}
\frac{N_A}{N_V}\to\frac{\Lambda_V^2}{\Lambda_A^2}\ .
\label{Cutoff}
\end{equation}
One should keep in mind that there is a certain amount of
arbitrariness in the exact relation between $N_V$ and $N_A$.
Clearly, physical results have to be \emph{independent} of this
arbitrariness in order to guarantee universality of physical
quantities. In other words, provided Eq.~(\ref{Cutoff}) is
satisfied, no additional information is needed for how the limit
$N_{V,A}\rightarrow \infty$ is taken in
Eqs.~(\ref{PiV},\ref{PiA}).  This point will be crucial in what
follows.

Equations (\ref{PiV}) and (\ref{PiA}) each contain sums over an
infinite number of states, and the expansion for large values of
$Q^2$ therefore is non-trivial. In particular, it does not
commute with the operation of summing over $n$. However, Ref.
\cite{BeanePRD} appears to argue that this is only true of the
individual sums in $\Pi_V$ and $\Pi_A$, because of the presence
of the parton model $\log{Q^2}$. Consequently, in the difference
$\Pi_{V}-\Pi_{A}$, which does not contain this logarithm, the
claim is that one recovers the OPE as a naive expansion in
inverse powers of $Q^2$, which commutes with the operation of
summing over $n$. According to Ref. \cite{BeanePRD}, one then
finds at large $Q^2$ (with $N_{V,A}\rightarrow \infty $)
\begin{eqnarray}\label{OPE}
  \Pi_{V}-\Pi_{A} &\sim &
  \frac{\left\{\sum_{n=1}^{N_V}F^2_V(n)-
  \sum_{n=1}^{N_A}F^2_A(n)\right\}}{Q^2}
  \qquad \qquad \qquad \qquad \qquad \nn \\
&+& \frac{\left\{\sum_{n=1}^{N_V}F^2_V(n)M_V^2(n)-
 \sum_{n=1}^{N_A}F^2_A(n)M_A^2(n)\right\}}{Q^4}
 \\ &+&\qquad  \mathcal{O}\left(\frac{1}{Q^6}\right) \nn\quad  .
\end{eqnarray}
In particular, then (again with $N_{V,A}\rightarrow \infty$),
\begin{eqnarray}\label{WSR1}
  \sum_{n=1}^{N_V}F^2_V(n)-
  \sum_{n=1}^{N_A}F^2_A(n)&=&0\ ,\qquad
  \mathrm{and}\ \\
  \sum_{n=1}^{N_V}F^2_V(n)M_V^2(n)-
  \sum_{n=1}^{N_A}F^2_A(n)M_A^2(n)&=&0 \label{WSR2}
  \ ,
\end{eqnarray}
from the absence of operators of dimension 2 and dimension 4 in
the OPE of $\Pi_{V}-\Pi_{A}$.  The claim of Ref. \cite{BeanePRD}
seems to be that these equations are well-defined,
regulator-independent large-$N_c$ QCD sum rules  and, thus, that
these sum rules can be used to restrict {\it ans\"atze} for the
hadronic spectra.

These claims are not justified. Eqs.~(\ref{WSR1},\ref{WSR2}) must
be supplied with information on how $N_{V,A}$ go to infinity,
because, as we shall see, the results actually depend on this.
Since the initial Eqs.~(\ref{PiV}, \ref{PiA}) are insensitive to
how this limit is taken (as long as Eq.~(\ref{Cutoff}) is
satisfied), this information can only be supplied by the hadronic
spectrum itself, which consequently makes
Eqs.~(\ref{WSR1},\ref{WSR2}) not very useful for constraining
that spectrum.

Of course all this happens because, in order for the operations
of summing over $n$ and expanding in $1/Q^2$ to commute, all sums
involved would need to be absolutely and uniformly convergent,
and this is clearly not the case. In particular, the absence of
the parton model logarithm cannot be a convincing reason for an
expansion such as Eq.~(\ref{OPE}) to make sense. The Adler
function, which is defined as
\begin{equation}\label{Adler}
  -Q^2 \frac{d}{dQ^2} \Pi_{V}(Q^2) \ ,
\end{equation}
by construction does not have a logarithm either, but it is easy
to convince oneself that it does not need to have an OPE with
numerators given by sums over resonance parameters
\cite{Golterman2,Shifman}.

As a matter of fact, there are at least three objections to the
claims of Ref.~\cite{BeanePRD}. First, the claim about chiral
symmetry restoration for highly excited mesons does \emph{not}
take place in the only situation where large-$N_c$ QCD is soluble
and everything is under good theoretical control, which is the
case of two dimensions \cite{Callan,Zhitnitsky}.%
\footnote{Note that two-dimensional large-$N_c$ QCD in the chiral
limit does exhibit spontaneous breaking of chiral symmetry, if
the chiral limit $m_{quark}\to 0$ is taken after the limit
$N_c\to\infty$. This is precisely the order of limits in which
't~Hooft solved this theory and does not contradict Coleman's
theorem. For details, see Ref.~\cite{Zhitnitsky} and Refs.
therein.}  There, one finds that the spectrum of meson masses,
which for highly excited states goes like
\begin{equation}\label{qcd2}
  M^2_{n} \sim \mathrm{constant}\times n\ ,
\end{equation}
actually \emph{alternates} in parity as one increases $n$ by one
unit. The conflict with the claims of Ref.~\cite{BeanePRD} can be
made most obvious if one considers scalar and pseudo-scalar
two-point functions $\Pi_{S,P}$. This is due to the fact that
$\Pi_S(Q^2)$ and $\Pi_P(Q^2)$ in two dimensions have
representations like those of $\Pi_{V,A}(Q^2)$ in the ordinary
four-dimensional case,\footnote{ There are several reasons why
$\Pi_{S,P}$ are actually the natural two-dimensional analogs of
$\Pi_{V,A}$ in four dimensions \cite{Einhorn}.} i.e.
Eqs.~(\ref{dispersion},\ref{Imvector},\ref{Imaxial},\ref{asympt}).
A rerun of the analysis of \cite{BeanePRD} in two dimensions
for $\Pi_{S,P}$ would conclude that the spectrum of excited
scalar and pseudo-scalar mesons should be degenerate; however,
clearly, it is not. Two-dimensional large-$N_c$ QCD is therefore
a counter example to the claim of Ref.~\cite{BeanePRD}.

Second, the functional dependence of Eq.~(\ref{OPE}) on $Q^2$ is
not correct. The currents in the correlator Eq.~(\ref{correlator})
have no anomalous dimensions.  The leading operator in its OPE
expansion is the quark condensate (squared) which has non-zero
anomalous dimensions even in the large-$N_c$ limit.  The
corresponding Wilson coefficient therefore has a residual
$\log{Q^2}$ dependence which is missing in Eq.~(\ref{OPE})
\cite{Chetyrkin}. Therefore Eq.~(\ref{OPE}) cannot be exact. This
was the point of view taken by the authors of Refs.~\cite{MHA} in
constructing an \emph{approximation} to large-$N_c$ QCD with a
\emph{finite} number of resonances. Clearly, in this case all
expansions at large $Q^2$ make definite sense \cite{Golterman2}.

If one considers Eq.~(\ref{OPE}) as some approximate statement, up
to such offending logs, then our third objection becomes relevant.
This has to do with the fact that these sums are ill-defined
because, unlike those in Ref. \cite{MHA}, they are infinite and,
as we already pointed out, must be regulated. This is why we
wrote Eqs.~(\ref{WSR1},\ref{WSR2}) with sums cut off at  $N_V$
and $N_A$. It is then important to remember that these sums
should not depend on the precise value taken for the cutoffs
$N_V$ and $N_A$ and, for instance, the same physical results
should be obtained choosing $N_V+a$ instead of  $N_V$ for
arbitrary finite $a$ in the limit $N_V\rightarrow\infty$. This is
a fundamental property which guarantees the universality of the
physics in any Quantum Field Theory and, in fact, one can readily
check that the original expressions of Eq.~(\ref{Euler}) satisfy
this requirement. However, the sum rules in
Eqs.~(\ref{WSR1},\ref{WSR2}) \emph{do} depend on $a$. Let us take
Eq.~(\ref{WSR1}), for example, and replace $N_V \rightarrow N_V +
a $ in Eq.~(\ref{WSR1}).  Using again the Euler-Maclaurin
summation formula one obtains:
\begin{eqnarray}\label{flaw}
  \sum_{n=1}^{N_V+a}F^2_V(n)-
  \sum_{n=1}^{N_A}F^2_A(n) &\approx&
  \int_0^{N_V+a-1}dn\ F^2_V(n)-
  \int_0^{N_A-1}dn\ F^2_A(n)+
  \cdots\nn \\
&\approx& F_V^2 (N_V+a-1)-F_A^2 (N_A-1)+ \cdots
\end{eqnarray}
where $F_{V,A}$ are given in Eq.~(\ref{asympt}) and the ellipsis
are terms subleading in $N_{V,A}$ as $N_{V,A}\rightarrow \infty$.
Use of Eqs.~(\ref{PT},\ref{Cutoff}) then shows the cancellation
of terms leading in the cutoff $\Lambda$ in the above expression,
but one also sees that the sum rule is $a$ dependent. This simply
means that this sum rule is not consistent with universality.

Let us study the issue of regulator dependence in some more
detail, using a recently analyzed {\it ansatz} for the vector and
axial-vector meson spectrum \cite{Golterman} as an example.  In
fact, a claim of Ref.~\cite{BeanePRD} is that the sum rules of
Eqs.~(\ref{WSR1},\ref{WSR2}) rule out this model. To illustrate
the point about regulator dependence made above, let us show how
one can choose to satisfy Eq.~(\ref{WSR1}), by tuning the cutoff
dependence in our model \cite{Golterman}, without any chiral
symmetry restoration at all.

In the model of Ref.~\cite{Golterman}, all $F_A(n)$ were taken
equal, $F_A(n)=F_A$, except for the pion, with $F_A(1)=F_\pi$, as
were all $F_V(n)=F_V$, except that of the lowest vector
resonance, the $\rho$, for which $F_V(1)=F_\rho$.  Masses were
taken to follow the Regge-like pattern \bea
M^2_V(1)&=&M^2_\rho\,,\label{masses}\\
M^2_V(n)&=&m_V^2+(n-2)\Lambda^2_V\,,\ \ \ n>1\,,\nonumber\\
M^2_A(1)&=&M_{\pi}^2=0\,,\nonumber\\
M^2_A(n)&=&m_A^2+(n-2)\Lambda^2_A\,,\ \ \ n>1\,,\nonumber \eea
where $m_{V,A}$ and $\Lambda_{V,A}$ are four parameters with
dimension of mass.  In Ref.~\cite{Golterman} various relations
generalizing the Weinberg sum rules were derived from the OPE
between the parameters of the model, $m_{V,A},\ \Lambda_{V,A},
F_{V,A},\ M_\rho,\ F_\rho$ and $F_\pi$. As we already emphasized,
there is a certain freedom in the choice of the exact relation
between the cutoffs $N_V$ and $N_A$, which can be made explicit
by choosing\footnote{In Ref.~\cite{Golterman} a specific choice
was made for these parameters, but it is straightforward to show
that none of the results obtained there depend on their values.}
\begin{equation}
\Lambda^2\equiv N_A\Lambda_A^2=N_V\Lambda_V^2+\mu^2,
\label{relation}
\end{equation}
where $\mu$ is a finite, but otherwise undetermined parameter.
The only requirement is that physically meaningful results should
not depend on $\mu$, and this is precisely where the sum rules in
Eqs.~(\ref{WSR1},\ref{WSR2}) run into trouble.\footnote{It is
important to recall that in the application of the OPE the scale
$\Lambda$ has to be taken much larger than any other scale in the
problem, including the (euclidean) momentum $Q$ flowing through
the current-current correlators.} Let us show this by applying
them to the {\it ansatz} of Ref.~\cite{Golterman}.  From
Eq.~(\ref{WSR1}), we obtain for this {\it ansatz} \be
F^2_\rho+(N_V-1)F_V^2-(N_A-1)F^2_A=F_\pi^2\,.\label{sransatz} \ee
Using Eqs.~(\ref{relation}) and (\ref{PT}) this can be rewritten
as \be \frac{N_c}{24
\pi^2}\;\mu^2=-F^2_\pi+F^2_A+F^2_\rho-F^2_V\,.\label{result} \ee
If $\mu$ satisfies this condition, Eq.~(\ref{WSR1}) is satisfied.
This demonstrates the problem with the sum rule Eq.~(\ref{WSR1}):
it depends on the choice of the values of the unphysical
parameter $\mu$ whether the sum rule will be obeyed or not.  The
divergences in both sums ({\it i.e.} the terms linear in $N_V$
and $N_A$ in Eq.~(\ref{sransatz})) do cancel as they should, but
there are ``left-over" finite terms which take on arbitrary
values depending on the details of the regularization.

Something similar also happens with the sum rule Eq.~(\ref{WSR2}).
Substituting our {\it ansatz} leads to {\it two} constraints on
the parameter $\mu$, one from the requirement that the linear
term in $\Lambda^2$ vanish, and one from the finite part. (The
leading terms, quadratic in $\Lambda^2$, cancel identically.)
These two constraints turn out to be incompatible with
Eq.~(\ref{result}), unless one requires the full vector and
axial-vector spectral functions to be equal, in which case one
finds that choosing $\mu=0$ satisfies all constraints (not
surprisingly).

This latter conclusion is precisely that advocated in
Ref.~\cite{BeanePRD}. However, our discussion makes it clear that
this is misleading: whether each individual sum rule is satisfied
or not depends on the detailed choice of the finite parts of the
regulator, represented here by the unphysical parameter $\mu$.
Consequently, the choice made in Ref. ~\cite{BeanePRD} is
arbitrary and ad hoc, not based on chiral symmetry, and not on
any other known property of QCD. What it does, in fact, is
arbitrarily enforcing the degeneracy of the spectrum, rather than
obtaining it as a result. In contrast, all sum rules derived in
Ref.~\cite{Golterman} are independent of the value of $\mu$. The
sum rules of Eqs.~(\ref{WSR1}) and (\ref{WSR2}) will only be
independent of the regulator if each of the sums in these
equations converges, and their derivation only holds if this is
the case.  The real spectrum of QCD is unlikely to satisfy this
requirement, in which case the sum rules cease to be meaningful.

We have already indicated why we believe that the derivation of
these sum rules from the OPE is flawed.  In addition to the
OPE-based argument, another derivation was given in
Ref.~\cite{BeaneJPG}.  There, chiral symmetry was implemented on
finite representations of the $SU(2)_L\times SU(2)_R$ algebra in
the null plane. However, it is clear that what is needed is the
implementation of chiral symmetry on infinite representations,
and one expects that consequently sums over states will again
have to be properly regularized.  This was not done in
Ref.~\cite{BeaneJPG}, making this derivation equally dubious.

It would of course be very interesting to infer knowledge about
the hadron spectrum from the OPE, as one does expect that the
higher-lying spectrum should be constrained by the OPE.
Unfortunately, in our opinion, all one can confidently say is
that the large-$k$ behavior of the coefficients $c_{k}$ in the OPE
\begin{equation}\label{opeope}
  \Pi_V(Q^2)- \Pi_A(Q^2) \sim \sum_k \frac{c_k}{Q^{2k}}\ ,
\end{equation}
is correlated with the spectrum of highly excited states
\cite{Zhitnitsky,Shifman,Golterman2}. Following Zhitnitsky
\cite{Zhitnitsky}, we write
\begin{eqnarray}\label{f}
  \Pi_V(Q^2)-\Pi_A(Q^2) &=&\sum_{n=1}^{\infty}
  \left(\frac{F^2_V(n)}{Q^2+M^2_V(n)}- \frac{F^2_A(n)}{Q^2+M^2_A(n)}
  \right)\nn \\
  &=&  \int_0^{\infty}\ \frac{dt}{e^t-1}\
  f(t,Q^2)\quad ,
\end{eqnarray}
where the function $f(t,Q^2)$ satisfies
\begin{equation}\label{ff}
  \frac{F^2_V(n)}{Q^2+M^2_V(n)}-
  \frac{F^2_A(n)}{Q^2+M^2_A(n)}=\int_0^{\infty} dt\  e^{-nt}
  f(t,Q^2)\ .
\end{equation}
This equation can be inverted with the help of the inverse
Laplace transform to read
\begin{equation}\label{ilaplace}
  f(t,Q^2)=\frac{1}{2 \pi i}\ \lim_{T\rightarrow \infty}\
  \int_{-iT}^{+iT}dn \ e^{nt}\ \left(\frac{F^2_V(n)}{Q^2+M^2_V(n)}-
  \frac{F^2_A(n)}{Q^2+M^2_A(n)}\right)\ .
\end{equation}
For the function $f$ to be uniquely determined, one needs to
specify the interpolation of the left-hand side of Eq.~(\ref{ff})
to non-integer values of $n$.  Here we have in mind that, at
least for large $n$, $M_{V,A}^2(n)=\Lambda_{V,A}^2n$ and
$F_{V,A}^2(n)=F_{V,A}^2$ for all positive $n$. The singularities
of the integrand lie to the left of the integration path, as they
should. Using that
\begin{equation}\label{Bernoulli}
  \frac{t}{e^{t}-1}=\sum_{k=0}^{\infty} \frac{B_k}{k!}\ t^k\quad ,\ \mathrm{with}
  \quad B_k \sim
  \frac{(2k)!}{2^{2k-1}\pi^{2k}}\ \left(1+\frac{1}{2^{2k}}+\cdots
  \right)\ ,
\end{equation}
where the $B_n$ are Bernoulli's numbers, one obtains the
asymptotic expansion
\begin{equation}\label{Piasymp}
  \Pi_V(Q^2)-\Pi_A(Q^2) \sim \sum_{k=0}^{\infty} \frac{B_k}{k!}
  \int_{0}^{\infty}dt\ t^{k-1} f(t,Q^2)\ .
\end{equation}

In order to calculate the integral in Eq.~(\ref{ilaplace}) one
completes the path with a semi-circular contour to the left of the
imaginary axis, and of radius $T$. The behavior of $f(t,Q^2)$ for
large values of $Q^2$ is dominated by the asymptotic behavior of
$F_{V,A}(n)$ and $M^2_{V,A}(n)$ for large $n$, and one finds
\begin{equation}\label{fasymp}
  f(t,Q^2)\ \sim \ \left(\frac{F_V^2}{\Lambda_V^2}
  \ e^{-\frac{Q^2}{\Lambda_V^2}\ t}-
   \frac{F_A^2}{\Lambda_A^2}\ e^{-\frac{Q^2}{\Lambda_A^2}\ t} \right)
\end{equation}
as $Q^2$ becomes large.\footnote{Note that the integral in
Eq.~(\ref{Piasymp}) converges for $k=0$ because of
Eq.~(\ref{PT}).} Inserting this into Eq.~(\ref{Piasymp}) one
finally obtains that, for large $k$, the coefficients of the OPE
in Eq.~(\ref{opeope}) behave like
\begin{equation}\label{c}
  c_k \ \sim \ \frac{(2k)!}{2^{2k-1}\ k \pi^{2k}}
  \left(F_V^2 \Lambda_V^{2(k-1)}- F_A^2 \Lambda_A^{2(k-1)}\right)\ .
\end{equation}
Equation (\ref{c}) expresses the difficulty. One would need to
know the asymptotic behavior of the coefficients $c_k$ in
large-$N_c$ QCD to be able to infer whether $\Lambda_V=\Lambda_A$
and $F_V=F_A$ or not, {\it i.e.} whether chiral symmetry is
restored for highly excited hadronic states.

In summary, the claims made in Refs.
\cite{BeanePRD,CohenIJMP,CohenPRD} to the effect that the OPE
constrains the spectrum of highly excited hadrons to be
degenerate parity eigenstates are unjustified. There may be some
indication \cite{Glozman} of chiral restoration in the current
experimental data on the spectrum of scalars and pseudoscalars
\cite{Anisovich}, but drawing any definite conclusions at this
stage seems premature. In our opinion, the question of chiral
symmetry  restoration for very highly excited hadrons is a very
interesting one which, however, remains completely open and
deserves further investigation.

\vskip 1 cm

\leftline{\bf Acknowledgments}

\vspace*{3mm}

We would like to thank Silas Beane, and S.P. would also like to
thank Marc Knecht and Eduardo de Rafael, for discussions. We would
also like to thank the Institute for Nuclear Theory at the
University of Washington (Seattle), where this work started, for
hospitality and support. MG was supported in part by the US Dept.
of Energy, and SP was supported by CICYT-AEN99-0766,
CICYT-FEDER-FPA2002-00748, 2001 SGR00188 and by TMR EC-Contracts
No. ERBFMRX-CT980169 (EuroDaphne) and HPRN-CT-2002-00311
(EURIDICE).

\end{document}